\documentclass[trackchanges]{aastex701}

\usepackage{amsmath}
\usepackage[nopatch]{microtype}
\usepackage{booktabs}
\usepackage{multirow}
\usepackage{xcolor}
\hypersetup{
    colorlinks=true,
    linkcolor=blue,
    citecolor=blue,
    urlcolor=blue
}

\begin{document}

\title{ 
Irregularity in Active Fast Radio Burst Repeaters and Magnetar Periodic Radio Pulses:\\ Time, Energy, and Frequency Analyses
}

\author[]{Ellen C.C. Lin}
\affiliation{Department of Electrophysics, National Yang Ming Chiao Tung University, Hsinchu, 300093, Taiwan}
\email[Ellen C.C. Lin]{jasonlin030203@gmail.com}

\author[]{Shotaro Yamasaki}
\affiliation{Department of Physics, National Chung Hsing University, 145 Xingda Rd., South Dist., Taichung 40227, Taiwan}
\email[Shotaro Yamasaki]{shotaro.s.yamasaki@gmail.com}

\author[]{Tomotsugu Goto}
\affiliation{Department of Physics, National Tsing Hua University, 101, Section 2. Kuang-Fu Road, Hsinchu, 30013, Taiwan}
\email[Tomotsugu Goto]{tomo@phys.nthu.edu.tw}

\author[]{Tetsuya Hashimoto}
\affiliation{Department of Physics, National Chung Hsing University, 145 Xingda Rd., South Dist., Taichung 40227, Taiwan}
\email[Tetsuya Hashimoto]{tetsuya@phys.nchu.edu.tw}

\begin{abstract}

Fast Radio Bursts (FRBs) are millisecond-duration radio pulses with largely unknown origins, with a subset exhibiting repeating behavior. Magnetars—highly magnetized neutron stars and a leading progenitor candidate for FRBs also produce similar but much fainter millisecond radio pulses, suggesting a possible connection between the two phenomena. The irregularity of the time series of repeating FRBs and magnetar pulses may provide insight into the underlying progenitor activity. In this study, we analyze time-series data from three 
repeating FRB sources (four datasets) and the Galactic magnetar SGR J1935+2154 to investigate potential patterns in burst arrival times, energy fluctuations, and peak-frequency shifts. We quantify the degree of randomness (Pincus Index; PI) and chaos (Largest Lyapunov Exponent; LLE) for these three parameters.  We find that waiting times across all repeating FRBs exhibit high PI (high randomness) and low LLE (low chaos), consistent with the behavior of magnetar radio pulses. This similarity suggests that both may share a common triggering mechanism. 
In contrast, the energy fluctuations of both repeating FRBs and magnetar pulses occupy the same region in PI–LLE phase space but display much larger scatter than the other two domains. We discuss the possibility that beaming effects or strong variability in radio-emission efficiency may explain their distinct behavior in the energy domain.

\end{abstract}




\section{Introduction} \label{sec:Intro}

Fast Radio Bursts (FRBs) are intense ($\sim$Jy) pulses of ($\sim$GHz) radio emission lasting only a few milliseconds from extragalactic space \citep{Lorimer07, Thornton13}. While a Galactic intrinsically-faint FRB-like event has been firmly linked to a magnetar SGR J1935+2154 \citep[e.g.,][]{andersen2020bright,bochenek2020fast,Hu2024}, the origins of cosmological FRBs remain uncertain. Although over 4,000 FRB sources have been discovered (\citealt{CHIME21}; CHIME Collaboration et al., submitted), the vast majority have been detected only once, which limits efforts to constrain their progenitors and emission mechanisms. In contrast, a subset of FRB sources are known to repeat, with tens of thousands of bursts detected from the most active ones when observed by highly sensitive telescopes such as FAST and Arecibo \citep[e.g.,][]{Li21, Jahns23, Xu22, Zhang23}.
A key open question in FRB studies is the intrinsic nature of their activity. In particular, it has been actively debated whether all FRBs are intrinsically repeaters \citep[e.g., ][]{Ravi19, James23, Yamasaki24}. Because active repeating sources produce numerous bursts, analyzing their rich datasets of burst time series could offer a unique opportunity to probe the underlying emission mechanism. 

Previous studies have mainly focused on the statistics of burst arrival times. A bimodal distribution of FRB waiting times, first reported by \citep{Li21}, was later found to be ubiquitous among active repeaters \citep[e.g., ][]{Zhang2023BurstProperties, Chen2022, Panda2025}. Two-point correlation analyses further revealed similarities to both earthquakes \citep{Totani23} and periodic magnetar radio pulses \citep{Tsuzuki24}, motivating models in which magnetar activity triggers repeating FRBs \citep{Luo25}.

Irregularity, or more specifically, chaos-randomness analysis, offers a complementary approach by directly quantifying the degree of stochasticity and dynamical stability in burst sequences. \citet{Zhang23} introduced the Pincus Index (PI) and the Largest Lyapunov Exponent (LLE) to characterize the dynamical behavior of astrophysical transients, comparing two repeating FRBs with stochastic processes such as Brownian motion. \citet{Yamasaki23} later applied this method to magnetar bursts, and \citet{Sang24} analyzed four repeating FRB sources in the time and energy domains. Nonetheless, existing applications remain limited either by the still-small number of repeating FRBs analyzed or by a focus restricted to the time and energy domains.

In this study, we build on these earlier works by applying the chaos–randomness analysis to a comparable set of repeating FRBs while broadening the parameter space examined. 
In particular, we incorporate the peak frequency domain, a key observable in major FRB emission models, such as the shock-powered scenario \citep[e.g., ][]{Metzger19} and magnetospheric scenario \citep[e.g., ][]{Lu20}, but one not previously explored using this method. 
Furthermore, for each source we extract the two largest burst series within continuous observing blocks, allowing us to test the internal consistency of the chaos–randomness properties across independent burst episodes from the same source. 

Finally, the Galactic FRB source magnetar SGR J1935+2145 has also exhibited periodic radio pulsations (PRPs) since 2020 \citep[e.g., ][]{Bailes21, Zhu23}. These PRPs are quasi-periodic—while they recur at a characteristic timescale corresponding to the neutron star’s rotation, their arrival times show noticeable deviations from strict periodicity \citep{Zhu23}, unlike the highly regular pulses of ordinary radio pulsars. 
Since such behavior in PRPs might be related to FRB activity \citep{Tsuzuki24, Kramer_2023}, we analyze these magnetar PRPs from SGR J1935+2145 within the same chaos–randomness framework to compare with repeating FRBs for the first time. 

This paper is organized as follows.
In \S\ref{s:data}, we describe the datasets used in our analysis.
In \S\ref{s:analysis}, we present the calculation methods and the resulting measurements for each dataset.
In \S\ref{s:Result}, we summarize the main findings, and in \S\ref{s:discussion}, we discuss their physical implications.
Finally, we conclude in \S\ref{s:conclusion}.

\section{Data}
\label{s:data}
In the analysis, we exploit four burst datasets from three active repeating FRB sources and one burst dataset of PRPs from a single magnetar source. All analyses use the barycentric arrival times as provided in the original data. 

{\it J23 \& J23g (FRB 20121102A)}:--The first dataset comprises 1,027 bursts (J23) from FRB 20121102A  detected with the 305-m Arecibo Telescope \citep{Jahns23}. The 849 events (J23g) are obtained by grouping sub-bursts according to the authors’ criteria. We mainly use the full burst sample (J23), while J23g is analyzed to test the impact of grouping.
The observation period of J23g and J23 is between 2018 October 18 and November 28. 
From these two datasets, we extract and analyze the burst energy and peak frequency to investigate. Their column names are ``\textit{Scaled energy / erg}'' and ``\textit{f\_cent / MHz}'' respectively. 

{\it Z23 (FRB 20220912A)}:--The second includes 1076 bursts from FRB 20220912A  (\citealt[hereafter Z23]{Zhang23}) detected by FAST with a precise location. The observation period of Z23 is between 2022 October 28 and December 22.
We extract the burst energy and peak frequency to investigate, and their column name are \textit{``Energy''} and \textit{``PeakFrequency''}. 

{\it Z22 (FRB 20201124A)}:--The third data set is taken from FRB 20201124A (\citealt{Zhou22}, hereafter Z22). Its observation period is between 2021 September 25 and 2021 September 28. The burst energy is not explicitly listed in the original data. Therefore, we calculate the burst energy  with

\begin{equation}
\label{eq:energy}
E = 10^{39} \,{\rm erg}\, \frac{4 \pi}{1+z}\left(\frac{D_{L}}{10^{28}\ {\rm cm}}\right)^{2}\left(\frac{F_{\nu}}{\rm Jy\cdot ms}\right)\left(\frac{\Delta \nu}{\rm GHz}\right) , \tag{1}
\end{equation}

where $D_{L}=450$ Mpc is the luminosity distance, $z=0.098$ is the redshift of host galaxy (\citealt{Xu22}), $F_{\nu}$ is the fluence, and $\Delta\nu$ is the bandwidth.
Therefore, we extract the fluence and bandwidth to derive burst energy.  Therefore, we import the column, \textit{``BWe  (MHz) ''}, \textit{`` F (mJy~ms) ''}, and \textit{`` nu0 (MHz) ''} in original dataset.

{\it X22 (FRB 20201124A)}:--The last repeating FRB dataset is again from FRB 20201124A (\citealt{Xu22}, hereafter X22), contains 1863 bursts observed by FAST. X22 report observations from 2021 April 03 to May 26.
Since the original paper did not provide peak frequency information, we adopt the 1D Gaussian–fitted peak frequencies from \cite{Yamasaki2025}, which were derived from the one-dimensional spectra extracted from the full dynamic spectra archived in \cite{Wang24}.
The burst energy is not explicitly provided in the original paper. We calculate it using equation (\ref{eq:energy}), following the approach of \cite{Xu22}, which is similar to the method described in Z22.
Therefore, we extract the fluence and signal bandwidth to derive burst energy. Their column names are \textit{``fluece  (Jy s)''} and \textit{``bandwidth of signal (MHz)''}.
 
{\it W24 (PRPs from SGR J1935+2154)}:--In addition to the FRB samples, we also investigate the PRPs from SGR J1935+2154. These PRPs are quasi-periodic—while they recur on a characteristic timescale corresponding to the neutron star’s rotation, their arrival times show noticeable deviations from strict periodicity, unlike the highly regular pulses of ordinary radio pulsars \cite{Zhu23}. In a previous study, \cite{Tsuzuki24} examined the relationship between these PRPs and repeating FRBs using correlation function analysis and found a notable similarity. Here, we aim to apply the chaos–randomness analysis to PRPs for the first time to enable a direct comparison with repeating FRBs.
This dataset consists of 563 periodic radio pulses from SGR J1935+2154 (\citealt{Wang24}, hereafter W24) observed with FAST between 2020 October 9 and October 30.
We use the burst energy and peak frequency directly from the original dataset, corresponding to the columns labeled \textit{``Energy (erg)''} and \textit{``Cntrl freq. (MHz)''}, respectively.

For the PRP source, although the full campaign spans 30 days, FAST was available for 17 of those days, with most sessions consisting of $\sim$2-hour continuous observing blocks, supplemented by two extended 6-hour blocks on MJDs 59140 and 59167. No bursts were detected after MJD 59152, even though additional observing blocks followed. For the FRB sample, observations typically consist of $\sim$1-hour continuous observing blocks per day during each observing campaign.
For both the FRB and PRP sources, we analyzed the two continuous observing blocks (corresponding to two observing days for these sources) with the highest and second-highest burst counts. This selection ensures that the bursts originate from the same active episode of each source, rather than mixing activity from separate and potentially unrelated intervals.

Each of these telescopes (FAST and Arecibo Telescope) has its own observational frequency band.
All datasets were observed in the L band: FAST covers 1000–1500 MHz, while the Arecibo Telescope spans 1150–1730 MHz.
Following Z23, we remove the outer 30 MHz of each band to avoid using events that may suffer from large biases in peak frequency due to limited-band effects. 

\section{Burst Irregularity Analysis}
\label{s:analysis}

 We analyze the irregularity in the FRBs' time sequence of waiting time, energy fluctuation, and peak frequency change. 
 Using arrival time $T_{i}$, burst energy $E_i$, and spectral peak frequency $f_{i}$ (each normalized by their mean value\footnote{Normalizing each quantity by its mean effectively removes redshift-dependent effects.}), we define the parameters: $\Delta T_{i} = T_{i+1}-T_{i} $, $\Delta E_{i} = E_{i+1}-E_{i} $, and $\Delta f_{i} = f_{i+1}-f_{i} $, which represents the differences between the $i$-th and $(i+1)$-th events.
 We then use the series $\Delta T_{i}$, $\Delta E_{i}$, and $\Delta f_{i}$ as the inputs for the analyses in \S ~\ref{sec:PI} (degree of randomness) and \S ~\ref{sec:LLE} (chaos). These calculations follow the methodology of \cite{Zhang24} and \cite{Yamasaki23}. We additionally analyze the peak frequency change, because it may play an important role in the radiation mechanism of FRBs. For each dataset, we compute the relevant quantities in the time, energy, and peak-frequency domains (details in \S ~\ref{sec:PI} and \S ~\ref{sec:LLE}).

\subsection{Pincus Index}
\label{sec:PI}
Pincus Index (PI) is to quantify the degree of randomness. Its calculation is based on the Approximate Entropy (ApEn) \citep{Pincus91}. Consider the original time series as $\textbf{x}=\{x_1, x_2,...,x_N\}$ with length \textit{N}. ApEn for $\textbf{x}$ is defined as following:

\begin{equation}
    \text{ApEn}(r,m;\textbf{x}) = \phi^{m}(r) - \phi^{m+1}(r), \tag{2}
\end{equation}

where

\begin{equation}
    \phi^m (r)=\frac{\sum_{i=1}^{N-m+1}\ln C_{i}^{m}(r)}{N-m+1}, \tag{3}   
\end{equation}

\begin{equation}
    C_{i}^{m}(r)=\frac{(\text{number of j such that $d[u_{i},u_{j}]<r$})}{N-m+1},
    \tag{4}
\end{equation}

where $\textbf{u}=
\{u_{i}\}_{i=1,...,(N-m+1)}$ and $u_{i}=\{x_{i},...,x_{i+m-1}\}$ is subsequence from initial series $\textbf{x}$ with length $m$, $d[u_{i},u_{j}]$ is Chebyshev distance, and $r$ is the distance threshold.
In our calculation, we set the threshold 
$r$ to vary over the interval $[0, std(\textbf{x})]$, where $std(\textbf{x})$ is the standard deviation of the \textbf{x}.
The Chebyshev distance is defined as the largest absolute difference between corresponding elements across the vectors. Low ApEn values imply the presence of patterns, indicating some level of predictability in the data series. In contrast, high ApEn values indicate high randomness.
However, only relying on ${\rm ApEn}_{\rm max}$, which is defined as the equation (\ref{eq:ApEnmax}) , for comparison with diverse phenomena is not robust. In order to resolve this condition, Z23 adopts the PI to remove interference by shuffling series elements in \textbf{x} as the equation (\ref{eq:PI})

\[
\label{eq:ApEnmax}
{\rm ApEn}_{\rm max}= \max\limits_{r}[{\rm ApEn}(\textbf{r})],
\tag{5}
\]

\[
\label{eq:PI}
{\rm PI} = \frac{{\rm ApEn}_{\rm max}(m;\textbf{x}_{\rm original})}{\langle {\rm ApEn}_{\rm max}(m;\textbf{x}_{\rm shuffled})\rangle}, 
\tag{6}
\]

and with given m and \textbf{x}. Here, we denote $\textbf{x}_{\rm original}$ as original data. The $\textbf{x}_{\rm shuffled}$ were generated by randomly shuffling the $\textbf{x}_{\rm original}$. We act random shuffling on the $\textbf{x}_{\rm original}$ 100 times, so there are 100 values of ${\rm ApEn}_{\rm max}(m;\textbf{x}_{\rm shuffled})$.
By dividing by ${\rm ApEn}_{\rm max}(m;\textbf{x}_{\rm shuffled})$, this normalization makes the series capable of comparing randomness with the other series. 

Throughout the analysis, we use $m=2$, following \cite{Zhang24} (we confirmed that the result is not significantly changed for $m>2$), so the subsequence \textbf{u} would be $\textbf{u}=\{u_{1},u_{2},...,u_{n}\}=\{[x_{1},x_{2}], [x_{2},x_{3}],...,[x_{N-1},x_{N}]\}$. The standard deviation value is the error bar for PI. In the calculation of the PI, its value would lie between 0 and 1. PI $=0$ means the series is completely predictable. On the other hand, PI $=1$ means the series is entirely random. We computed PI values for $\Delta T_{i}$, $\Delta E_{i}$, and $\Delta f_{i}$. 

\begin{table*}
\centering
\caption{\label{tab: Table 1}
PI values of waiting time, energy difference, and peak frequency fluctuation for each dataset. The number of events used ($N_{\rm event}$) and the corresponding observation date are also listed. Short horizontal lines in the FRB portion separate different sources. We used the following datasets of repeating FRBs: Z23 \citep{Zhang2023BurstProperties}, J23 and J23g \citep{Jahns23}, X22 \citep{Xu22}, and Z22 \citep{Zhang22}, as well as the PRP dataset W24 \citep{Wang24}. }
\begin{tabular}{l l c c c c c}
\hline
Transient & Dataset &$N_{\rm event}$&
Observation date (MJD) &
Waiting time & Energy difference & Peak frequency fluctuation\\\hline\hline
\multirow{10}{*}{FRB} 
& Z23-1 &277 &59882 &$0.96\pm 0.03$  & $0.80\pm 0.02$  & $0.89\pm 0.02$ \\
& Z23-2 &193 &59880 &$1.01\pm 0.03$  & $0.88\pm 0.03$  & $0.94\pm 0.03$ \\  \cline{2-7}
& J23g-1&190 &58439 &$0.99\pm 0.03$  & $0.90\pm 0.03$  & $0.94\pm 0.02$ \\
& J23g-2&178 &58432 &$1.02\pm 0.03$  & $0.88\pm 0.02$  & $0.98\pm 0.03$ \\ 
& J23-1 &221 &58439 &$0.99\pm 0.03$  & $0.94\pm 0.03$  & $0.98\pm 0.03$ \\
& J23-2 &215 &58432 &$1.02\pm 0.03$  & $0.98\pm 0.03$  & $0.98\pm 0.03$ \\ \cline{2-7}
& X22-1 &60  &59315 &$1.07\pm 0.06$  & $0.68\pm 0.05$  & $0.84\pm 0.08$ \\
& X22-2 &55  &59314 &$1.05\pm 0.06$  & $0.72\pm 0.05$  & $1.01\pm 0.07$ \\ 
& Z22-1 &632 &59485 &$1.01\pm 0.02$  & $0.83\pm 0.01$  & $0.95\pm 0.01$ \\
& Z22-2 &292 &59484 &$1.12\pm 0.04$  & $0.87\pm 0.02$  & $0.99\pm 0.09$ \\ \hline
\multirow{2}{*}{PRP}  & W24-1 &139 &59132 &$0.97\pm 0.04$  & $0.89\pm 0.03$  & $1.05\pm 0.06$ \\
& W24-2 &137 &59133 &$0.96\pm 0.04$  & $0.84\pm 0.04$  & $0.98\pm 0.05$ \\\hline
\end{tabular}
\end{table*}

\label{sec:LLE}
\subsection{Largest Lyapunov Exponent}
\label{sec:LLE}
The Lyapunov Exponent (LE) quantifies how rapidly two nearby trajectories diverge in phase space, and therefore measures a system’s sensitivity to initial conditions. A positive LE indicates exponential divergence and is the hallmark of chaotic dynamics \citep{Wolf1985}. Because a system is classified as chaotic if at least one LE is positive, we use the Largest Lyapunov Exponent (LLE) to characterize the degree of chaos in our burst sequences, following \citet{Zhang24}.
We use NonLinear measures for Dynamical Systems  (\texttt{nolds}), a Python-based module which provides the algorithm of \cite{Eckmann86} (\texttt{nolds.lyap\_e}) to approach the LLE. In our calculations, we set all parameter setting is similar to \cite{Yamasaki23} and \cite{Zhang24}. Moreover, we find that LLE is almost independent of the embedding dimension \citep{Sang24}. The one difference from PI that the LLE is the maximum value of the whole LE spectrum by the Python module \texttt{nolds} that we use. In practice, defining a meaningful uncertainty for the LLE is inherently difficult because it depends sensitively on the chosen finite length and noise properties of the time series. The module itself returns a single deterministic value; therefore, we treat the LLE as exact for the purpose of this analysis \citep[e.g.,][]{Zhang24,Yamasaki23}. We computed three LLE values for $\Delta T_{i}$,$\Delta E_{i}$, and $\Delta f_{i}$ respectively.
\begin{table*}
\caption{\label{tab: Table 2} Same as Table \ref{tab: Table 1}, but for LLE values.}
\centering
\begin{tabular}{l l c c c c c}
\hline
Transient & Dataset & $N_{\rm event}$&
Observation date (MJD) & Waiting time & Energy difference & Peak frequency fluctuation\\\hline\hline
\multirow{10}{*}{FRB} 
& Z23-1  & 277 & 59882 & $0.080$  & $0.123$  & $0.051$ \\
& Z23-2  & 193 & 59880 & $0.049$  & $0.109$  & $0.052$ \\  \cline{2-7}
& J23g-1 & 190 & 58439 & $0.077$  & $0.176$  & $0.043$ \\
& J23g-2 & 178 & 58432 & $0.077$  & $0.231$  & $0.063$ \\ 
& J23-1  & 221 & 58439 & $0.087$  & $0.229$  & $0.045$ \\
& J23-2  & 215 & 58432 & $0.052$  & $0.322$  & $0.049$ \\ \cline{2-7}
& X22-1  & 60 & 59315 & $0.051$  & $0.226$  & $0.007$ \\
& X22-2  & 55 & 59314 & $0.036$  & $0.105$  & $-0.004$ \\ 
& Z22-1  & 632 & 59485 & $0.261$  & $0.111$  & $0.085$ \\
& Z22-2  & 292 & 59484 & $0.695$  & $0.138$  & $0.025$ \\ \hline
\multirow{2}{*}{PRP}  & W24-1  & 139 & 59132 & $0.031$  & $0.068$  & $0.089$ \\
& W24-2 & 172 & 59133 & $0.085$  & $0.038$  & $0.010$ \\\hline
\end{tabular}

\end{table*}

\section{Results}
\label{s:Result}
\begin{figure*}
  \centering
  \includegraphics[width=0.8\textwidth]{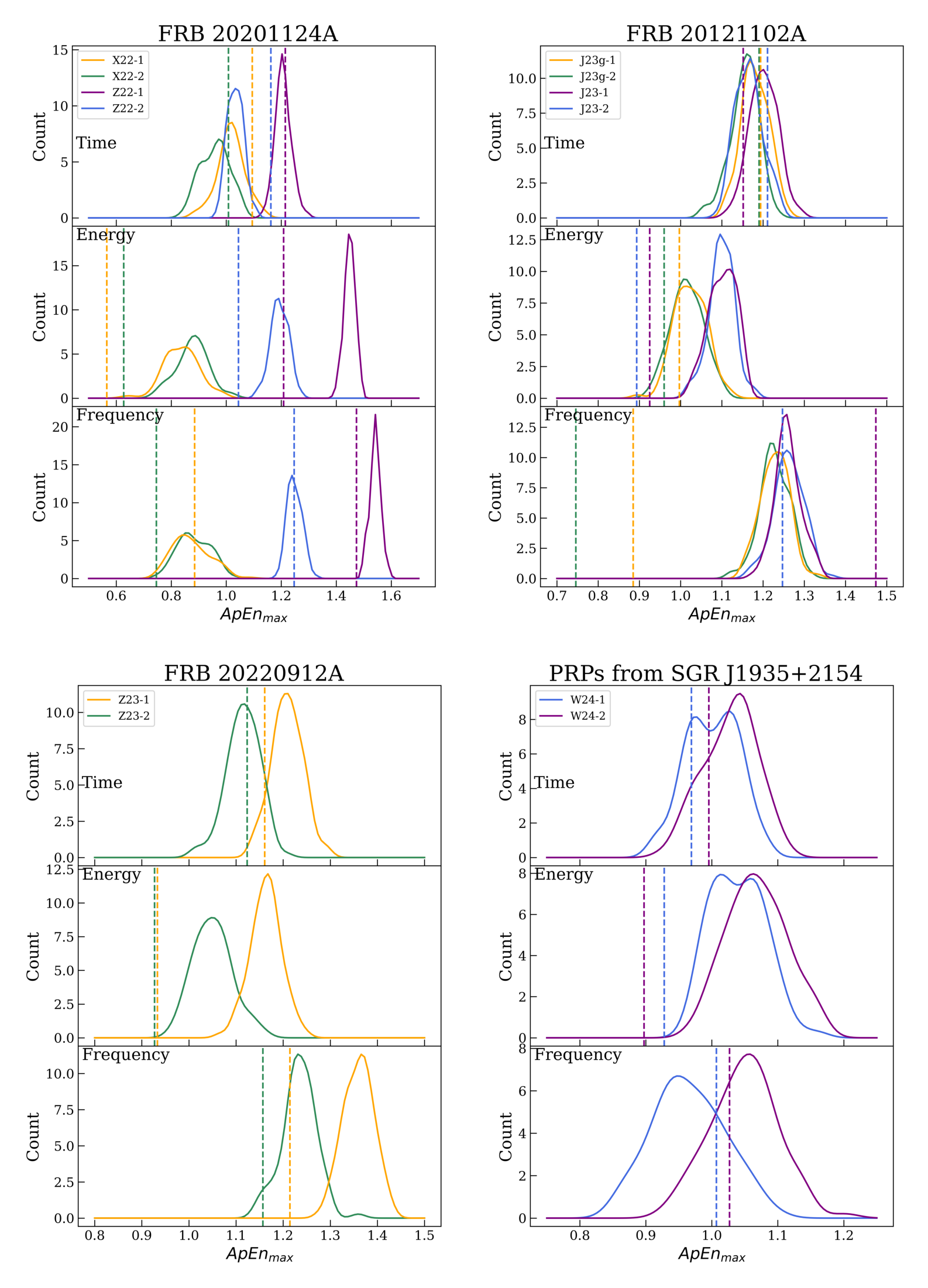}
  \caption{\label{fig: Figure 1.}The ${\rm ApEn}_{\rm max}$ distribution for FRB 20201124A, FRB 20121102, FRB 20220912A, and PRPs from SGR J1935+2154. Kernel density estimates show the distributions of ${\rm ApEn}_{\rm max}$ computed from 100 shuffled realizations of each series. The dashed lines indicate the ${\rm ApEn}_{\rm max}$ values of the original series, with each dashed line matching the color of its corresponding solid line.
  }
\end{figure*}

We first show the PI result in \autoref{tab: Table 1}, and then the LLE result in \autoref{tab: Table 2}. 
First, we focus on PI value and ${\rm ApEn}_{\rm max}$ value. 
\autoref{fig: Figure 1.} shows the ${\rm ApEn}_{\rm max}$ value for the original series marking as dash-line and the the kernel density estimate plots are the distribution of ${\rm ApEn}_{\rm max}$ for the 100 times shuffled series. 
\autoref{tab: Table 1} list the PI value and their standard deviation (the error bar).
As we can see from \autoref{fig: Figure 1.}, all the waiting time series of repeating FRBs adhere to a random organization, with the mean PI value of $1.02\pm 0.03$.  
The mean PI value is calculated across all repeating FRB datasets (X22, Z22, Z23, J23g, and J23), and the reported mean value and its uncertainty ($1.03 \pm 0.03$) correspond to the average and standard deviation of these dataset means.
We also confirm that J23 and J23g produce nearly identical PI and LLE values, showing that the sub-burst grouping does not affect the degree of randomness or chaos for FRB 20121102A.
However, the mean PI value in the Energy domain is $0.84\pm 0.08$. 
The larger standard deviation than the others reflects that there are some events which are close to a random system (such as J23g-1 and -2) and some far from a random system. In the frequency domain, the distributions of PI values are broader than the distributions in the Time domain but less dispersed than in the Energy domain. 

\begin{figure*}
  \centering
  \includegraphics[width=0.9\textwidth]{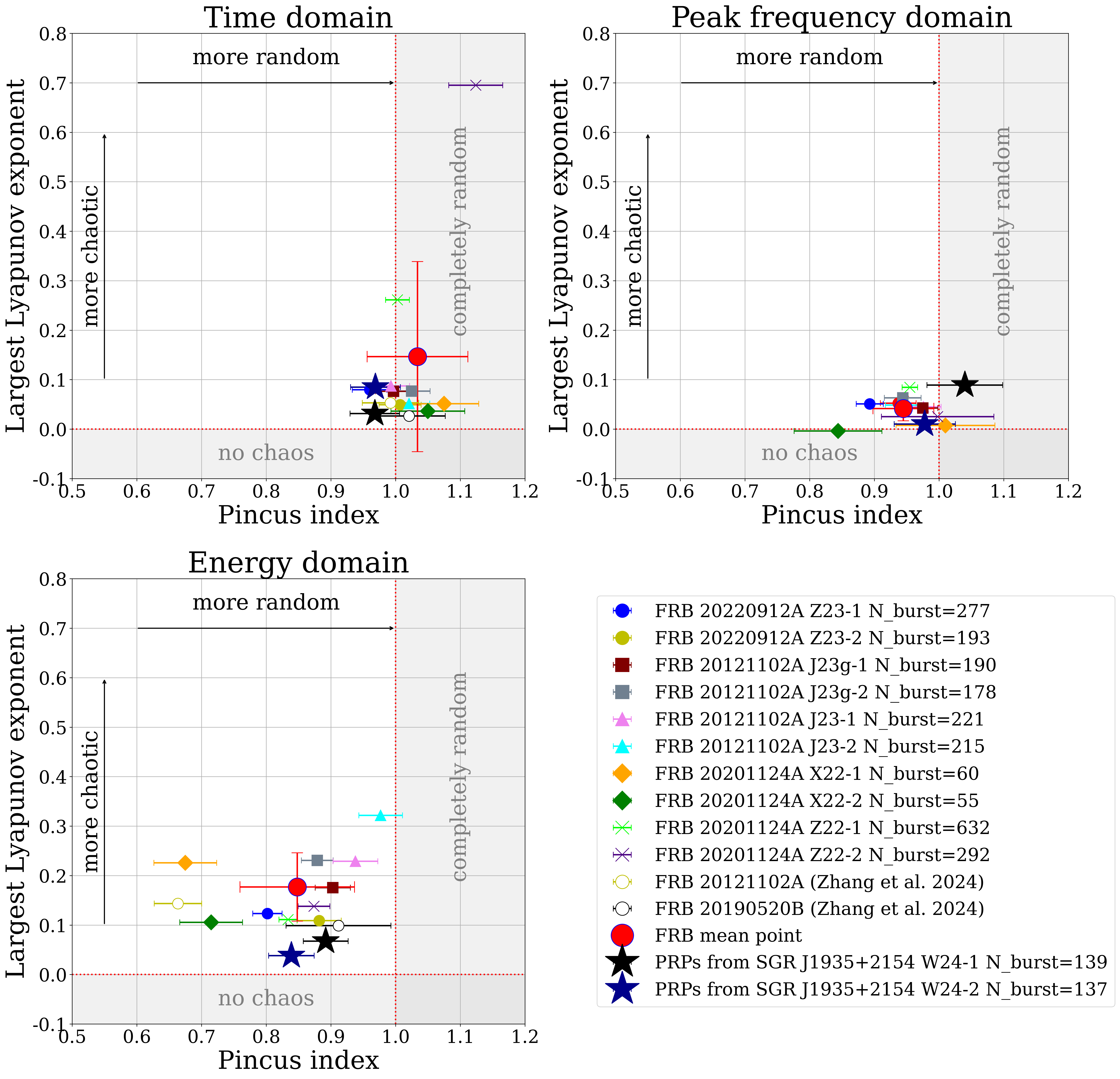}
  \caption{\label{fig: Figure 2.} 
  Chaos–randomness plane: Pincus Index versus Largest Lyapunov Exponent for repeating FRBs and radio pulses, shown for the waiting time (top left), peak frequency (top right), and energy (bottom left) domains. Data points for FRB 20121102A and FRB 20190520B are adapted from \cite{Zhang24}. The “FRB mean point” represents the average Pincus Index and LLE computed from all FRB datasets analyzed in this work (excluding the two sources from \citealt{Zhang24}). Light grey regions indicate extreme conditions, representing either complete randomness or the absence of chaos. } 
  \label{fig:example}
\end{figure*}

According to \autoref{fig: Figure 2.} and \autoref{tab: Table 2}, we find that most of the repeating FRBs and the PRPs tend to show low chaos in the time domain. However, Z22-1 and Z22-2 significantly deviate from that of the majority of events, which show a noticeably higher degree of chaos (0.261 and 0.695). The physical reason for this behavior is yet to be understood. We require further investigation to fully understand it. We now pay our attention to the energy domain, all repeating FRBs exhibit a wider spread along the degree of chaos axis (the standard deviation of FRB mean point is 0.069) than the two other parameters. The PRPs datasets (W24) display the least degree of chaos. We discuss the comparison with each domain and previous works is in \S \ref{s:discussion}. We focus on the last part, the frequency domain. Their the LLE value are clustered in the range from $-0.01$ to 0.01. Moreover, the PRPs' behavior in the randomness-chaos phase space does not manifest a significant difference. 

\section{Discussion}
\label{s:discussion}
\autoref{fig: Figure 2.} shows the chaos-randomness phase space, illustrating a comparison with each repeating FRB events and the radio pulses of SGR J1935+2154.
In the time domain, a clear clustering appears in the region characterized by high randomness and low chaos.. 
The clustering pattern suggests that the arrival times of the repeating FRBs and PRPs share similar behavior in the chaos-randomness phase space, irrespective of progenitor.
Another scenario is that the arrival time is only weakly correlated with their progenitor.
The PRPs show a similar trend with repeating FRBs, both populating the high-randomness and low-chaos region. 
This behavior is also consistent with the analysis of repeating FRBs reported in \cite{Sang24}. 
The result in the time domain explains that the Radio pulses and the repeating FRB have common properties in the time domain.

In the energy domain, we find a clear separation between data points on the phase-space diagram. 
Moreover, the standard deviation of PI in the energy domain is ~0.078, which is larger than those in the time domain (0.034) and peak-frequency domain (0.062). 
\cite{Yamasaki24} reported that PI values are uniformly distributed across repeating FRBs and magnetar X-ray bursts.
Our result is consistent with their finding.
We compare the repeating FRBs with the PRPs from SGR J1935+2154. 
The comparison shows repeating FRBs and Radio pulses have a similar PI value (the mean value of repeating FRBs is 0.84, and Radio pulses are 0.89 and 0.84 in the Energy domain).
However, their LLE values differ: the mean LLE of repeating FRBs is 0.177, whereas the two PRP events yield 0.031 and 0.085.
These differences suggest that the burst energy fluctuations of the repeating FRBs is different from those of the PRPs. 
It implies that there are certain mechanisms that affect the PI and LLE of the burst energy. 
We speculate that different emission mechanisms may have varying energy transition rates from total energy to burst energy, leading to variations in the distribution and stability of the burst energy.

In the peak frequency domain, we apply this methodology for the first time.
The distribution in the randomness-chaos space of the peak frequency is akin to waiting time. The PI value is lower than in the time domain (average PI value in the time domain: 1.01 and in the peak frequency domain: 0.95). On the other hand, the distribution of the LLE in the frequency domain is similar to the time domain. So, it may imply that their behavior on the time series is similar. 

Overall, we find that FRBs and PRPs from a magnetar SGR J1935+215 show similarities in both the time domain and the peak frequency domain, whereas the energy domain exhibits larger scatter. One possible reason for the variability of burst energies on the chaos–randomness plane is that FRB emissions are likely beamed as those from pulsars \cite{Connor2020}. We only observe a fraction of the beam, so the intrinsic energy could differ from the measured value. Another possibility is the non-uniform conversion efficiency of rotational or magnetic energy into radio emission. For pulsars, it is known that the radio conversion efficiency can vary, with some sources exhibiting efficiencies of $\sim10^{-2}$--$10^{-4}$ relative to the spin-down luminosity \cite{Szary2014}. Consequently, the burst energy time series can show different randomness–chaos properties even within the same source or across different sources.

On the other hand, burst arrival times of FRBs, which are related to the intrinsic trigger mechanisms \citep{Totani23,Tsuzuki24,Luo2025}, and peak frequencies, which are related to the plasma physics at the emission site \citep{Lyu2024,Yamasaki2025}, may better trace the intrinsic source properties. Our finding that FRBs and PRPs share similar properties only in the time and peak frequency domains suggests that the underlying emission physics of these phenomena could be similar.

\section{Conclusions}
\label{s:conclusion}
In this work, we investigated the irregularity—quantified by chaos and randomness—of repeating FRBs by analyzing a dataset comparable to that of \cite{Sang24} and performing a direct comparison with the PRP sample. We extended the chaos–randomness analysis to include the peak-frequency domain, which has not been explored in previous studies.
In the time domain, both repeating FRBs and PRPs exhibit high randomness and low chaos, consistent with magnetar radio pulses. In the peak-frequency domain, they display a similar behavior, reinforcing the idea that the time and spectral characteristics may share a common emission mechanism. In contrast, in the energy domain, repeating FRBs and PRPs occupy the same general region in chaos–randomness space but show much larger scatter, suggesting that variability in radio-emission efficiency or beaming effects may influence their energy distribution.
The differences observed in the energy domain may reflect how total energy is converted into radio emission, potentially varying across different progenitor types. Overall, these findings highlight both the commonality in the underlying emission mechanism and the possible diversity of repeating FRB progenitors.
Future investigations could include a more extensive collection of repeating FRBs and other potential transient events. By adding datasets of repeating FRBs, we could examine whether repeating FRBs form subclasses or share similarities with other events. Moreover, this analysis could be extended to additional physical quantities if future models or observational evidence identify new parameters that play a key role in the mechanism of repeating FRBs \citep[e.g., ][]{Hu2023}.

\begin{acknowledgments}
ECCL thanks Tomoki Wada and the NTHU/NCHU cosmology group for their helpful comments. ECCL acknowledges the Taiwan Astronomical Observatory Alliance (TAOvA) and its summer student internship program for partial financial support of this research.
SY acknowledges the support from the National Science and Technology Council of Taiwan (NSTC) through grant numbers 113-2112-M-005-007-MY3 and 113-2811-M-005-006-. 
TH acknowledges the support from the NSTC through grants 113-2112-M-005-009-MY3, 113-2123-M-001-008-, 111-2112-M-005-018-MY3, and the Ministry of Education of Taiwan through a grant 113RD109. 
TG acknowledges the support of the NSTC through grants 110-2112-M-005-013-MY3.
\end{acknowledgments}

%




\bibliography{reference}{}
\bibliographystyle{aasjournalv7}



\end{document}